\documentclass[aps,twocolumn,showpacs,pre,UTF8]{revtex4}
\usepackage{amsmath,epsfig,amssymb,subfigure,bm,dsfont}
\usepackage{amsmath}
\usepackage{txfonts}
\usepackage[titletoc]{appendix}
\usepackage{amssymb}
\usepackage{array}
\usepackage{mathrsfs}
\usepackage{subfigure,overpic}
\usepackage{dcolumn}
\usepackage{graphicx}
\usepackage{epstopdf}
\usepackage{float}
\usepackage{flushend}
\usepackage{balance}
\usepackage{lipsum}
\usepackage[colorlinks,
            linkcolor=blue,
            anchorcolor=blue,
            citecolor=blue]{hyperref}

\begin{document}

\title{Indirect magnetic signals mediated by a single surface band in Weyl  semimetals}
\author{Hou-Jian Duan}
\email{dhjphd@163.com}
\affiliation{Guangdong Provincial Key Laboratory of Quantum Engineering and Quantum
Materials, School of Physics and Telecommunication Engineering, South China Normal
University, Guangzhou 510006, China}
\author{Yong-Jia Wu}
\author{Ming-Xun Deng}
\author{Ruiqiang Wang}
\email{wangruiqiang@m.scnu.edu.cn}
\author{Mou Yang}
\affiliation{Guangdong Provincial Key Laboratory of Quantum Engineering and Quantum
Materials, School of Physics and Telecommunication Engineering, South China Normal
University, Guangzhou 510006, China}

\begin{abstract}
Recently, abundant transport phenomena characterizing the surface states of Weyl semimetals (WSMs) have
been reported. To generate these phenomena, electrons have to complete a closed intersurface orbit. Due to the unavoidable impurities in real materials, this orbit would be destroyed by the impurity scattering, which limits the detection of the surface states in WSMs. Here, we investigate the RKKY interaction between magnetic impurities, solely mediated by a single surface band, in semi-infinite WSMs. It is found that peculiar oscillations and slowly decaying laws of the RKKY interaction can act as the signals to capture the dispersive nature of the surface states of WSMs. The underlying physics is attributed to two effects: the band-edge effect and the bending effect of the surface band, which can control the RKKY interaction individually or compete with each other to produce more complex magnetic behaviors. In addition, the band-edge effect together with the finite Fermi energy would result in another interesting oscillation with battering pattern. All the results are significantly different from that in previous literatures where surface states have to couple with bulk states (or other surface states of different spins) to generate nonzero magnetic interaction. Compared to the previous models of surface states, the model here is more practical and is helpful for the deeper understanding of the surface magnetic properties in WSMs.

\end{abstract}

\maketitle


\section{introduction}
Weyl semimetals (WSMs), as the firstly discovered three-dimensional topological semimetals, have attracted extensively attention due to their peculiar electronic structures and potential applications in spintronics. Different from topological insulators whose topology is protected by a considerable energy gap, the bulk band of WSMs is gapless but still remains the topological nature\cite{WSM1}. For the simplest model of WSM, the corresponding topological property is carried by a pair of Weyl points with opposite chiralities\cite{Okugawa}. Each Weyl point corresponds to a magnetic or an anti-magnetic
monopole, which supports topological charge $1$ or $-1$ and can be characterized by the nonzero Berry curvature\cite{berry1,berry2,berry3}. Due to the crystalline symmetry, the Weyl points can only be created or annihilated in pairs. According to the bulk-surface correspondence, topologically protected surface states would arise as a WSM with finite size is considered\cite{Wieder,Mou}. Different from the closed Fermi surface of topological insulators, the surface states of WSMs on the Fermi surface form open Fermi arcs connecting the Weyl points.
\par
To realize the WSM phase, one can split the degenerated Dirac points into two Weyl points by breaking the time-reversal symmetry\cite{floquet1,floquet2,doping1} or the inversion symmetry\cite{Okugawa}. For example, WSMs with broken time-reversal symmetry can be obtained by applying a beam of off-resonant light in nodal-line semimetals (NLSMs)\cite{floquet1} or Dirac semimetals (DSMs)\cite{floquet2}. Alternatively, similar WSMs can also be realized by doping magnetic impurities in DSMs\cite{doping1}.  In addition, various magnetic materials (${\rm HgCr_2Se_4}$\cite{Weng0}, ${\rm Y_2Ir_2O_7}$\cite{Vishwanath}, ${\rm Co_3Sn_2S_2}$\cite{Kumar}, and ${\rm Co_2}$-based Heusler compounds\cite{Felser,Vergniory}) have been proposed as the candidates for WSMs. Specially, ${\rm Co_3Sn_2S_2}$ has already been established as a magnetic WSM by using the angle-resolved photoemission spectroscopy\cite{D.F.}. So far, WSMs with broken inversion symmetry are mainly focused on the noncentrosymmetric transition-metal monosphides\cite{Weng}, including TaP\cite{SYXu,NXu}, NbP\cite{NbP1}, TaAs\cite{TaAs1,TaAs2,TaAs3}, and NbAs\cite{NbAs1}.
\par
The verification of WSMs in experiments have further prompted researchers' interests on the physical properties of WSMs. One of the most intriguing topic is how to detect the surface states of WSMs. To deal with this problem, many literatures have studied the surface states-related transport properties in WSMs and various phenomena are revealed, e.g., strong Friedel oscillations on the surface of WSMs\cite{Friedel}, anomalous quantum oscillations contributed by the Fermi arcs\cite{quantum1,quantum2,quantum3}, nonlocal dc voltage and sharp resonances in the transmission of electromagnetic waves induced by the unique intersurface cyclotron orbits\cite{cyclotron orbits}, peculiar magnetic-field-dependent magnetoconductivity\cite{magnetoconductivity}, three-dimensional quantum Hall effect\cite{Hall}, unusual magnetothermal transport\cite{magnetothermal}. From these nontrivial phenomena, transport signatures can be extracted for characterizing the surface states of WSMs. Noting that the above phenomena are induced by the unique intersurface orbit, i.e., electrons should be transported from one Fermi arc (on one surface) to another (on the opposite surface) via bulk Weyl monopoles. However, this orbit could be destroyed by the impurity scattering since defects or impurities are unavoidable in real materials. Moreover, to allow electrons to complete this orbit before scattering
off an impurity, a WSM with proper thickness $L$ is required, i.e., $L\ll l$ ($l$ denotes for the mean-free path). This limits the detection of the surface states in WSMs. Therefore, it is highly desirable to find another way to identify the surface states of WSMs, especially for the single surface band in semi-infinite (or large thickness $L$) WSMs without considering the intersurface process.
\par
The RKKY interaction between magnetic impurities has attracted our attention since it is sensitive to the behaviors of the electrons near the Fermi surface, and thus can be used to characterize the dispersive nature of materials\cite{graphene1,graphene2,graphene3,graphene4,graphene5,Weyl1,Weyl2,Weyl3,topo1,surface1,surface2,surface3,surface4,surface5,NLSM,alpha,semiDirac}. Naturally, the RKKY interaction is expected to be available for the single surface band of WSMs. So far, two literatures have already discussed the RKKY interaction on the surface of WSMs\cite{surface4,surface5}, it is found that the surface states-induced RKKY interaction survives only when the surface states couple with bulk states (or other surface states of different spins). In other words, no magnetic signals arise if a single surface band is considered in WSMs. To solve this problem, we propose two mechanisms as origins of the nonzero surface states-contributed RKKY interaction in WSMs, namely, the band-edge effect and the bending effect of the surface band. For different mechanisms, significantly different RKKY behaviors (slowly decaying laws and peculiar oscillations) would be generated. Furthermore, we have explored the competition of the above two effects, which results in more complex RKKY behaviors. In addition, the case of finite Fermi energy is discussed. From these discussions, various magnetic signals are extracted for characterizing the dispersive nature of the single surface band in WSMs.
\par
The paper is organized as follows: In Sec. II, a
minimal model of the WSMs is introduced, and the method to calculate the surface contribution of the RKKY interaction is presented. In Sec. III, the two
mechanisms of the nonzero surface contribution are analyzed respectively. The slowly decaying laws, as well as the peculiar oscillations, are also exhibited. Further more, we have explored the effect of the competition of the two mechanisms on the surface contribution. Additionally, the cases of finite Fermi energy are discussed. Finally, a short summary is
given.

\section{Model and Method}
\begin{figure}[th]
\centering \includegraphics[width=0.35\textwidth]{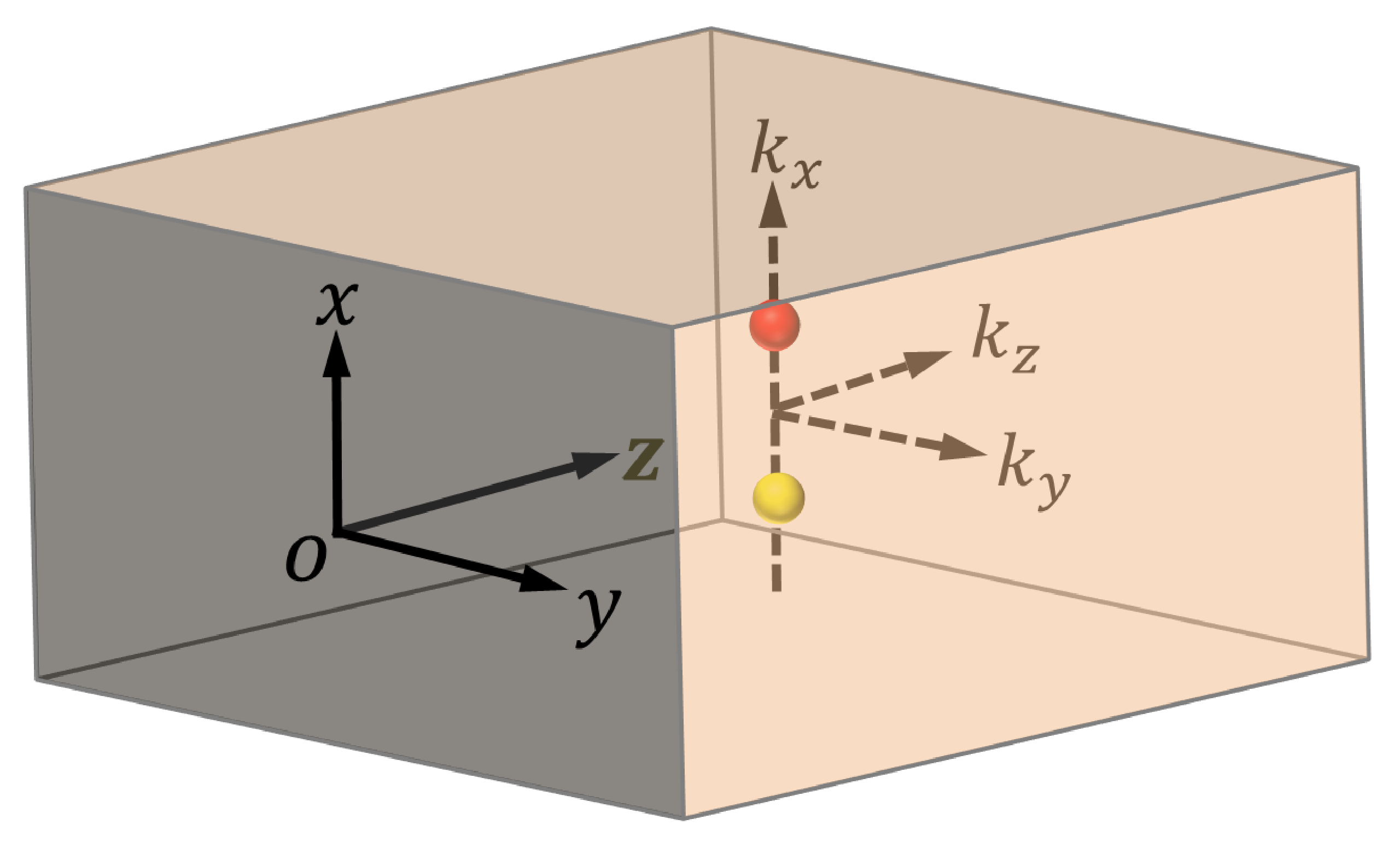}
\caption{(Color online) Schematic diagram for a WSM with two splitting Weyl nodes along the $k_x$ direction. The surface (the shadow face) is at $z = 0$ and the model here is infinite in other two ($x$ and $y$) directions.}
\end{figure}
We start with a minimal model of WSM whose low-energy Hamiltonian is given by
\begin{equation}
H_{_{\rm WSM}}=v\left(k_\parallel^2-k_0^2\right)\tau_z+v_zk_z\tau_y+\gamma k_y\tau_x+v_0k_\parallel^2\tau_0,
\end{equation}
where $k_\parallel^2=k_x^2+k_y^2$, and $\tau_i$ refers to the Pauli matrix operating in the orbital space. The first three terms describe a WSM with the band energy $\epsilon_{\pm}=\pm[v^2(k_\parallel^2-k_0^2)^2+v^2_zk_z^2+\gamma ^2 k_y^2]^{1/2}$. The distinctive feature for the bulk band $\epsilon_{\pm}$ is the two Weyl nodes, which carry opposite chiralities and are located at the positions of $(\pm k_0,0,0)$. Due to the nontrivial band topology, there exists surface band whose projection at zero Fermi energy is a straight line connecting two Weyl nodes. This straight line is bent by the last term of Eq. (1), i.e, changed to be an arc (known as Fermi arc). The model of Eq. (1) can be realized by considering the effect of a periodic driving to the NLSMs\cite{floquet1}, a detailed derivation is given in the Appendix I.
\par
To consider the effect of the surface states on the RKKY interaction, a semi-infinite WSM is studied. As shown in Fig. 1, the WSM is placed in the right half-plane ($z > 0$) and the other half ($z< 0$) is assumed to be a vacuum. The surface is at $z = 0$, and the model here is infinite in other two ($x$ and $y$) directions. Noting that the momentum $k_z$ here is not a good
quantum number, which  have to be replaced by the operator $k_z=-i\partial_z$. An incident wave $Ce^{-ik_z z}$ is assumed to injected along $z$ direction. Since the wave of the surface states is mainly bound to the $x$-$y$ surface, the solution of $k_z$ becomes imaginary. By considering the continuity conditions of the boundary between left and right regions, the wave functions and the energy band of surface states at the surface ($z = 0$) can be solved as,
\begin{eqnarray}
\begin{split}
  E\left(k_\parallel<k_0,v_0\right)&=v_0k_\parallel^2+\gamma k_y, \\
\Psi\left(k_\parallel<k_0,v_0,\mathbf{r}\right)&=\sqrt{\zeta} e^{i\mathbf{k_\parallel}\mathbf{r}}\left(
                                                                    \begin{array}{c}
                                                                      1 \\
                                                                      1 \\                                                                  \end{array}                                                              \right),
\end{split}
\end{eqnarray}
where $\zeta=v(k_0^2-k_\parallel^2)/v_z$. Noting that $E$ and $\Psi$ vanish for the momentum out of the circle $k_\parallel=k_0$. Substituting $k_\parallel$ with $|k_x|$ in the limitation $k_\parallel<k_0$ of $E$ and $\Psi$ (as well as in $\zeta$), the band $E(|k_x|<k_0,v_0=0)$ recovers to the case discussed in previous literatures\cite{surface4,surface5}, where the higher-order momentum $v_0k_{\parallel}^2$ is ignored and the dispersion is infinite in $k_y$ axis, as shown in Fig. 2(b).
\par
\begin{figure}[th]
\centering \includegraphics[width=0.45\textwidth]{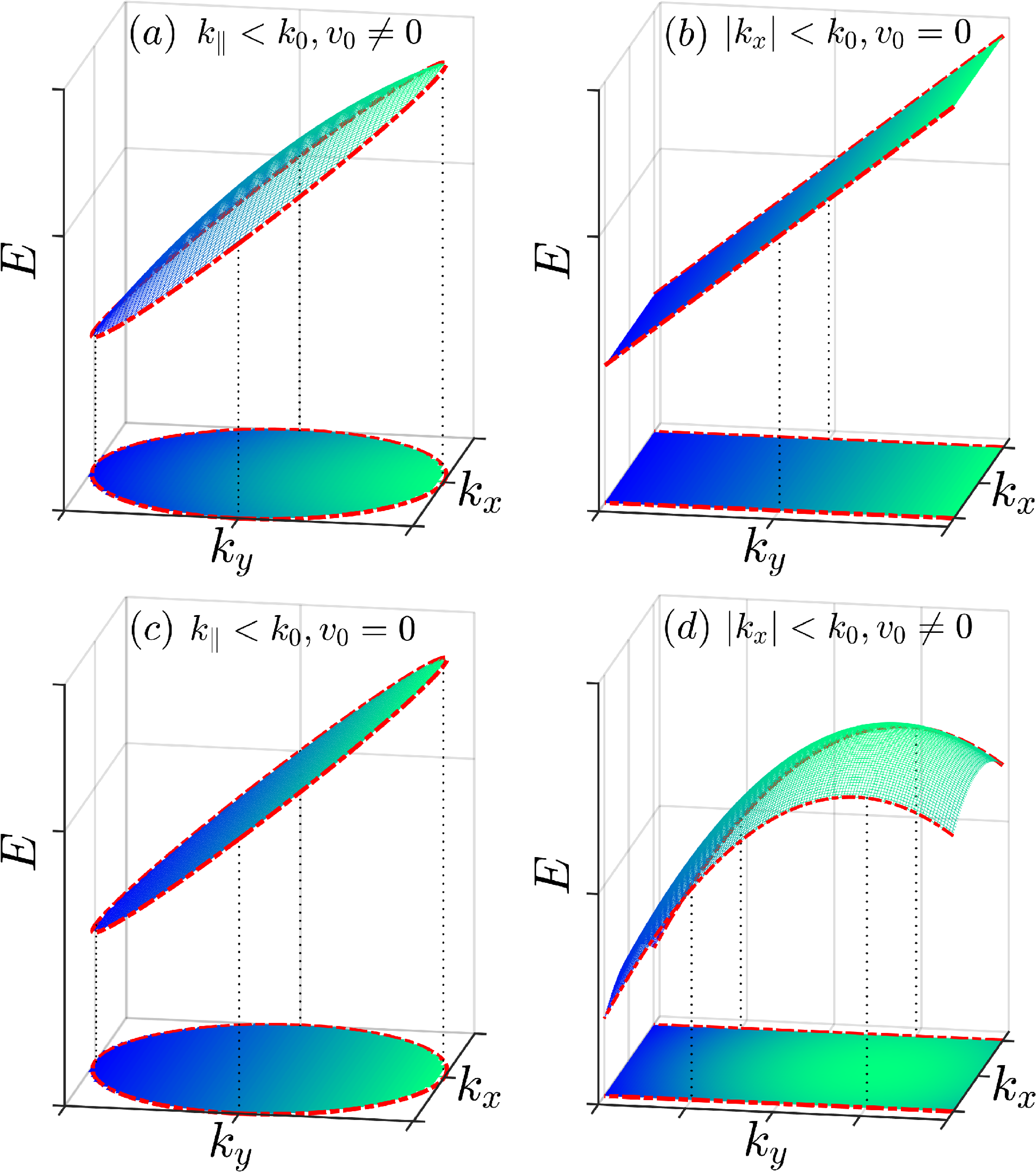}
\caption{(Color online) The dispersion of the surface band $E$ and its projection on the $k_x$-$k_y$ plane with (a) $k_\parallel<k_0,v_0\neq0$, (b) $|k_x|<k_0,v_0=0$, (c) $k_\parallel<k_0,v_0=0$ and (d) $|k_x|<k_0,v_0\neq0$. The red dashed lines denote for the band edges.}
\end{figure}
The dispersion $E(k_\parallel<k_0,v_0\neq0)$ is plotted in Fig. 2(a), where the red dashed lines refer to the band edges. The shape of the dispersion here is significantly different from that of $E(|k_x|<k_0,v_0=0)$ [Fig. 2(a)]. To figure out the detailed differences, one can consider the following two cases, respectively. (1) By simplify setting $v_0=0$, the surface band is simplified as $E(k_\parallel<k_0,v_0=0)=\gamma k_y$ , as shown in Fig. 2(c). Here, all surface states are confined to a circle ($k_\parallel=k_0$), which makes the surface band look like a tilted disc. This is different from that in Fig. 2(b) where the dispersion is ribbon-like, i.e., finite in $k_x$ axis but infinite in $k_y$ axis.  (2) By changing the limitation $k_\parallel<k_0$ to $|k_x|<k_0$ and setting $v_0\neq0$, the energy of the surface band is rewritten as $E(|k_x|<k_0,v_0\neq0)=v_0\left(k_\parallel^2-k_0^2\right)+\gamma k_y$. Compared to the band in Fig. 2(b), the only difference is that the band is bent by the term $v_0\left(k_\parallel^2-k_0^2\right)$, as shown in Fig. 2(d). In next chapters, we will show that the above two cases can contribute nonzero RKKY interaction individually. Moreover, more complex RKKY behaviors would be generated when the two cases coexist.
\par
Two magnetic impurities are assumed to be placed on the surface ($z=0$) of the WSM with the positions ${\mathbf{r}_1}$ and ${\mathbf{r}_2}$.  Considering the spin-exchange interaction  ($s$-$d$ model) between impurities and host electrons, the system Hamiltonian $H_{0}$ is rewritten as
\begin{equation}
H=H_{0}+H_{int}=H_{0}-J_0\sum_{i=1,2}\mathbf{S}_i\cdot \mathbf{s}_i,
\end{equation}
where $J_0$ stands for the strength of the exchange interaction, $\mathbf{S}_i$ is the spin of impurity at site $i$, and $\mathbf{s}_i=\frac{1}{2}c^\dag_{i\alpha}\sigma_{\alpha\beta}c_{i\beta}$ refers to the spin of host electrons with $\sigma_{\alpha\beta}$ being the matrix element of the Pauli operator in real spin space. Mediated by the itinerant host electrons, an indirect exchange interaction (i.e., RKKY interaction) between two impurities is generated. Considering the case of weak coupling $J_0$ between impurities and electrons, $H_{int}$ can be regarded as a perturbation. Using the perturbation theory by keeping exchange interaction $J_0$ to the second-order term\cite{rkky1,rkky2,rkky3,rkky4}, the effective coupling between the magnetic impurities can be written as
\begin{equation}
H_{_{RKKY}}=-\frac{\lambda ^{2}}{\pi }\mathrm{Im}\int_{-\infty }^{u_{F}}\mathrm{%
Tr}[\left( \mathrm{\mathbf{S}_{1}}\cdot \mathbf{\sigma }\right) G\left(
\omega,\mathbf{R} \right) \left( \mathrm{\mathbf{S}_{2}}\cdot \mathbf{\sigma
}\right) G\left( \omega,-\mathbf{R} \right) ]d\omega ,
\end{equation}%
where $\mathbf{R}=\mathbf{r}_1-\mathbf{r}_2$, $u_{F}$ is the Fermi energy and $G\left( \pm\omega,\mathbf{R}\right) $ is the retarded Green's function with respect to $H_{0}$ in real space.
\par
To calculate the RKKY interaction, the retarded Green's function of real space has to be derived. In the Lehmann's representation\cite{surface1,surface2},  $G\left( \omega ,\mathbf{R}\right)$ can be constructed by using the energy $E$ and the corresponding wave functions $\Psi$ of Eq. (2), given by
\begin{eqnarray}
\begin{split}
G\left(\omega,\mathbf{R}\right) =&\sum_{\mathbf{k}}\frac{\Psi
\left(k_\parallel<k_0,v_0,\mathbf{r}_{1}\right) \Psi ^{\dag }\left(k_\parallel<k_0,v_0, \mathbf{r}%
_{2}\right) }{\omega -E+i\eta }, \\
=&g\left(\omega,\mathbf{R}\right)\left(\tau_0+\tau_x\right),
\end{split}
\end{eqnarray}%
with
\begin{eqnarray}
\begin{split}
g\left( \omega ,\pm\mathbf{R}\right)=\sum_{\mathbf{k}}\zeta e^{\pm i\mathbf{kR}}/\left(\omega -E+i\eta\right).
\end{split}
\end{eqnarray}%
Plugging the above equations into the Eq. (4) and tracing the spin and orbital degrees of freedom, the RKKY interaction can be rewritten in the form of $H_{_{RKKY}}=J\mathbf{S}_{1}\cdot\mathbf{S}_{2}$ with $J$ reads as
\begin{equation}
J=-\frac{8\lambda^2 }{\pi }\mathrm{Im}\int_{-\infty }^{u_{F}}[ g\left(
\mathbf{R},\omega \right)  g\left( -\mathbf{R},\omega \right) ]d\omega ,
\end{equation}
For the sake of simplicity, we still regard $g(\omega,\mathbf{R})$ as the Green's function since it preserves the main properties of  $G( \omega ,\mathbf{R})$. Here, we only focus on the RKKY interaction contributed by the surface states since the surface contributions (impurities in $y$ axis) decay much more slowly than the bulk contributions and would play a leading role in the long range (i.e., large impurity distance).

\section{Results and Discussion}

\subsection{Mechanisms of nonzero RKKY interaction by a single surface band in WMSs}
Physically, the Green's function $g(\omega,\mathbf{R})$ ($\mathbf{R}=\mathbf{r}_1-\mathbf{r}_2$) describes the scattering process of electrons from one impurity ($\mathbf{S}_1$ with position $\mathbf{r}_1$) to another ($\mathbf{S_2}$ with position $\mathbf{r}_2$), and $g(\omega,-\mathbf{R})$ corresponds to the inverse process. Thus, before showing the nonzero surface contribution, one can study the Green's functions to capture the related mechanisms.
\par
First, we review the Green's function $g(\omega,\pm\mathbf{R})$ of the surface band $E(|k_x|<k_0,v_0=0)$ [Fig. 2(b)], which has been explored in previous literatures\cite{surface4,surface5}. For impurities placed in the $y$ axis, $g(\omega,\pm R_y)$ reads as
\begin{eqnarray}
g\left(\omega,\pm R_y\right)=\frac{2vk_0^3}{3\pi v_z\gamma}ie^{\pm iR_y\omega/\gamma}\Theta\left(\mp R_y\right).
\end{eqnarray}
Here, $\Theta(x)$ is the Heaviside step function, and we set $R_y>0$ for simplicity. According to the above equation, one can see that electrons of surface states can be scattered from $\mathbf{r}_2$ to $\mathbf{r}_1$ by the nonzero Green's function $g(\omega,- R_y)$, while the inverse process can not be realized due to the vanished $g(\omega,R_y)$. In other words, the trajectory of the electrons are always open, which naturally does not generate effective RKKY interaction\cite{roundtrip}. The case is different when impurities are placed in the $x$ axis, where the Green's function is solved as
\begin{eqnarray}
g\left(\omega,\pm R_x\right)=i\frac{v}{\pi v_z\gamma}\frac{{\rm sin}\left(k_0R_x\right)-k_0R_x{\rm cos}\left(k_0R_x\right)}{R_x^3}.
\end{eqnarray}
Although the nonzero $g\left(\omega,R_x\right)g\left(\omega,-R_x\right)$ constructs a closed loop for the itinerate electrons, no effective interaction arises. The reason is that $g\left(\omega,R_x\right)g\left(\omega,-R_x\right)$ here is a real number while the interaction [Eq. (7)] is determined by the imaginary part of $g\left(\omega,\mathbf{R}\right)g\left(\omega,-\mathbf{R}\right)$.
\par
\begin{figure}[th]
\centering \includegraphics[width=0.49\textwidth]{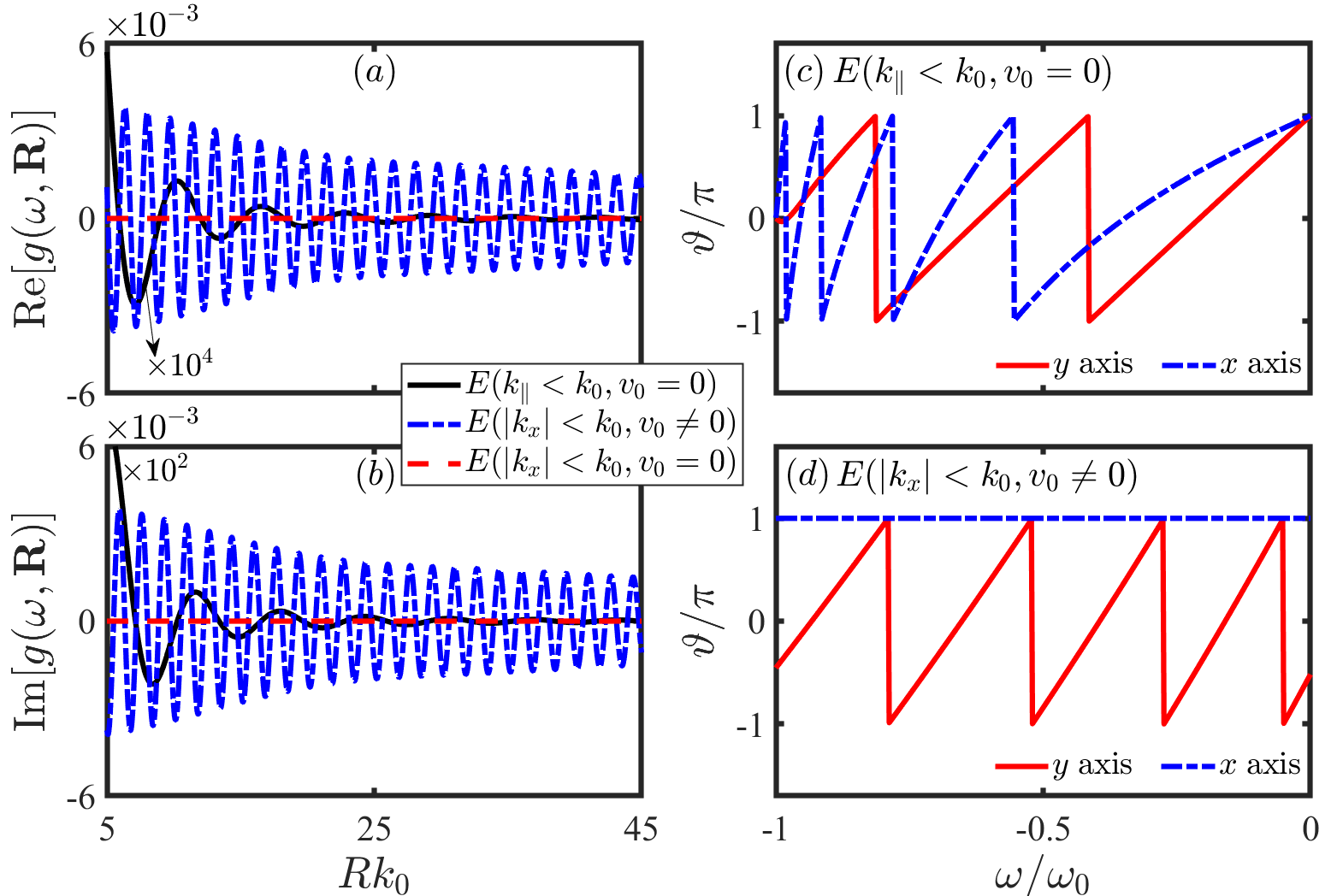}
\caption{(Color online) (a,b) Spatial dependence of the Green's function $g\left(\omega,\mathbf{R}\right)$ with $\omega=-0.01\omega_0$ and $\mathbf{R}=(0,R_y)$. The dependence of the angle $\vartheta$ on the energy $\omega$ in the cases of (c) $E(k_\parallel<k_0,v_0=0)$ and (d) $E(|k_x|<k_0,v_0\neq0)$ with $Rk_0=15$. The solid (dashed) lines refer to the case with impurities placed along $y$ ($x$) axis. Here, we set $\gamma=-0.3v_z$, $\hbar\omega=0.12{\rm eV}$, $\omega_0=-\gamma k_0$, and other parameters $v=4.34{\rm eV \AA^2}$, $v_z=2.5{\rm eV \AA}$, $\kappa_0=0.206 {\rm \AA}^{-1}$, $v_0=-0.993{\rm eV{\AA}^2}$(if $v_0\neq0$) are extracted from ${\rm Ca_3P_2}$\cite{Chan} material.}
\end{figure}
Taking into account the above scenarios, the mechanism for generating the nonzero interaction can be summarized in two steps. Firstly, electrons should display a round trip between impurities. Secondly, a nontrivial phase factor $e^{i\vartheta}$ ($\vartheta\neq 0,\pi$) has to be accumulated after the round trip. $e^{i\vartheta}$ is defined as
\begin{eqnarray}
e^{i\vartheta}=\frac{g\left(\omega,\mathbf{R}\right)g\left(\omega,-\mathbf{R}\right)}{|g\left(\omega,\mathbf{R}\right)g\left(\omega,-\mathbf{R}\right)|}.
\end{eqnarray}
Here, we set $-\pi<\vartheta\leq\pi$. $\vartheta=\pi$ corresponds to the case of  Eq. (9). The first condition can be easily met by the modified dispersions in Figs. 2(c-d). To confirm this point, one can check the amplitude of the Green's function. Noting that $g(\omega,-\mathbf{R})$ is always nonzero for arbitrary values of $v_0$ (or for arbitrary limitations of $k_{x,\parallel}$). Thus, we only plot the Green's function $g(\omega,\mathbf{R})$, which bears full responsibility for the open trajectory of the electrons [Eq. (8)]. For comparison, the case of $E(|k_x|<0,v_0=0)$ is also plotted. As shown in Figs. 3(a) and 3(b), original vanished Green's function (red lines) is changed to be a finite one (black and blue lines) by either changing the limitation of $|k_x|<k_0$ to $k_\parallel<k_0$ or turning on the parameter $v_0$. Thus, nonzero $g(\omega,\mathbf{R})$ together with $g(\omega,-\mathbf{R})$ would provide electrons a round trip between impurities. For $E(k_\parallel<k_0,v_0=0)$, the corresponding nonzero $g(\omega,\mathbf{R})$ is a result of the band-edge effect, i.e., the band edge (dashed line in Fig. 2(c)) acts like a wall and allows the electrons near the band edge to be backscattered. This explanation would be further verified by our analytical results in Sec. III-B. To understand the nonzero $g(\omega,\mathbf{R})$ of $E(|k_x|<k_0,v_0\neq0)$, one can simply check the Fermi velocity. Due to the bending effect (Fig. 2(d)) of the surface band, the original negative Fermi velocity $v_{y}=\gamma$ can be changed to be a positive one $v_{y}=\gamma+2v_0k_y$ (if $k_y<-\gamma/2v_0$), which also allows the backscattering behavior for electrons (i.e., nonzero $g(\omega,\mathbf{R})$ arises). To check whether the second condition is satisfied, we plot $\vartheta$ in Figs. 3(c) and 3(d). Obviously, almost all energies $\omega$ would contribute nontrivial phases (i.e., $\vartheta\neq 0,\pi$) expect for the case of $E(|k_x|<k_0,v_0\neq0)$ with impurities placed in $x$ axis. This means that nonzero RKKY interaction can be realized by either of the above effects, which is further verified in the next subsections.
\par
 In the following subsections, we only focus on the RKKY interaction with impurities placed in the $y$ axis since the slowly-decaying law occurs in this configuration. Compared to this configuration, the fast-decaying interaction with impurities in the $x$ axis is insignificant, and is shown in the Appendix II.

\subsection{RKKY interactions mediated by $E(k_\parallel<k_0,v_0=0)$ }
\begin{figure}[th]
\centering \includegraphics[width=0.45\textwidth]{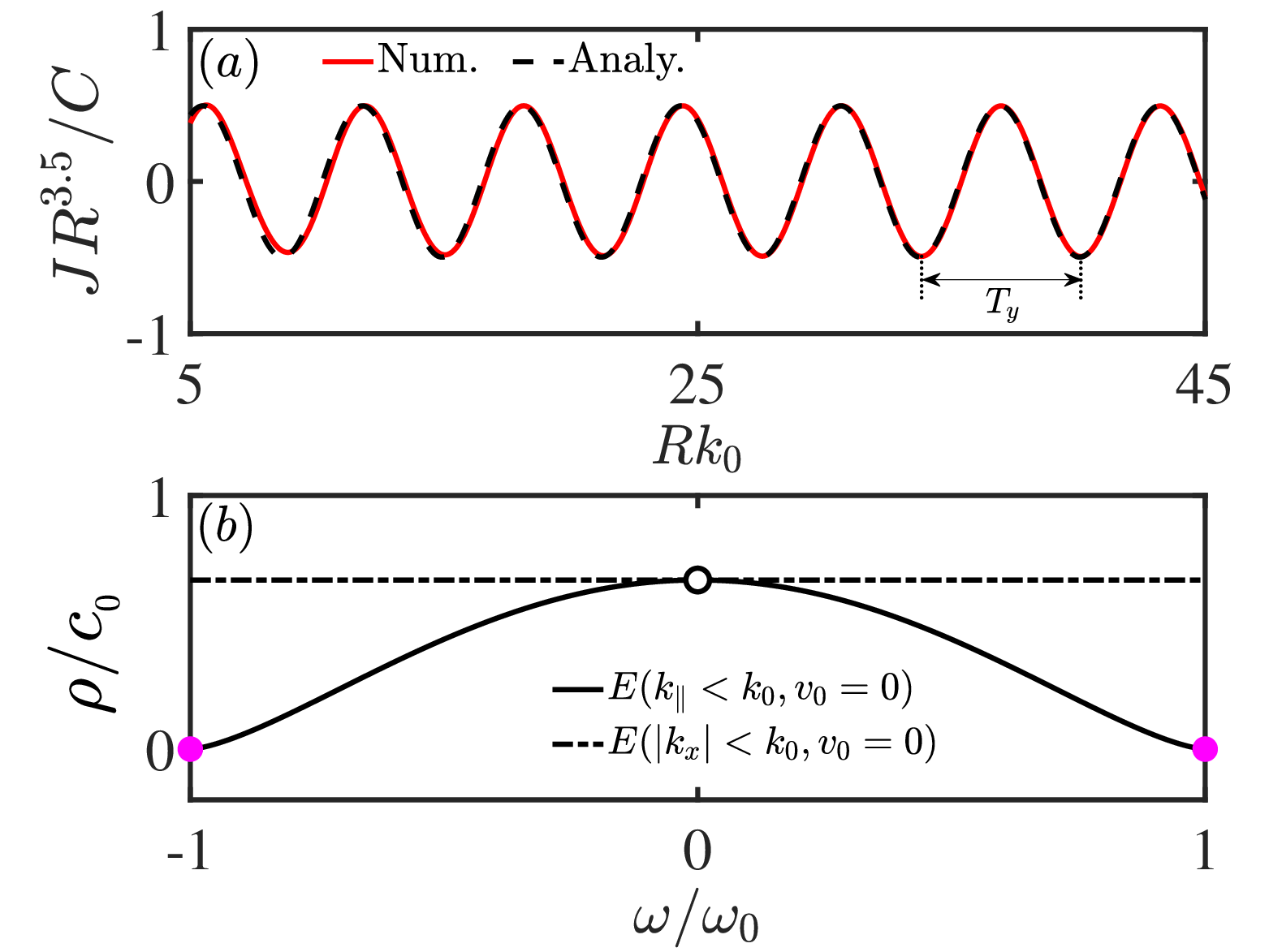}
\caption{(Color online) (a) The $R$-dependent RKKY interaction contributed by $E(k_\parallel<k_0,v_0=0)$ with impurities placed in $y$ axis. The solid line denotes for the numerical result calculated from Eqs. (6,7) and the dash line refers to the analytical result from Eq. (12). $C=\lambda^2/(2\pi)^3$ and other parameters are the same as that in Fig. 3(c). (b) DOS $\rho\left(\omega\right)$ with $c_0=v k_0^3/(-\pi^2\gamma v_z)$ for the cases of $E(k_\parallel<k_0,v_0=0)$ and $E(|k_x|<k_0,v_0=0)$.}
\end{figure}
In this chapter, we focus on the RKKY interaction of zero Fermi energy ($u_F=0$). Plugging the function $g(\omega,\pm\mathbf{R})$ of Eq. (6) into the Eq. (7), the RKKY interaction contributed by $E(k_\parallel<k_0,v_0=0)$ can be calculated numerically, as plotted in Fig. 4(a). For impurities deposited along $y$ axis, the interaction falls off as $1/R^{7/2}$ with the impurity distance $R$, which decays much more slowly than that of the bulk contributions ($1/R^5$) of WSMs\cite{Weyl1,Weyl2,Weyl3}. More interestingly, the interaction $J$ exhibits a peculiar oscillation, whose period $T_y=2\pi/k_0$ is determined by the distance $k_0$ between the band edge and the center $\mathbf{k_\parallel}=0$ of the band. This oscillation is quite unexpected since the Fermi wave number here is $k_F=u_F/\gamma=0$ and no separated Weyl points are located on $k_y$ axis. Noting that all oscillations of the RKKY interaction in previous works are usually induced by the finite Fermi wave number $k_F\neq 0$\cite{graphene1,graphene2,graphene3,graphene4,graphene5,Weyl1,Weyl2,Weyl3,topo1,surface1,surface2,surface3,surface4,surface5,NLSM,alpha,semiDirac} or the splitting of the Weyl/Dirac points\cite{graphene1,graphene2,graphene3,graphene4,graphene5,Weyl1,Weyl2,Weyl3,surface4,surface5,NLSM,alpha}. The underlying physics of the peculiar RKKY behavior here is attributed to the band-edge effect [i.e., the scattering behaviors (from $\mathbf{r_1}$ to $\mathbf{r_2}$) of the electrons near the band edge, as stated in Sec. III-A], which not only affects the decaying law but also determines the period of the oscillation. By detecting the period $T_y=2\pi/k_0$, the position of the band edge in momentum space can be accurately identified.
\par
To further understand the band-edge effect and the resulting RKKY behaviors [Fig. 4(a)], we analyze the density of  states (DOSs) in Fig. 4(b), where the cases of $E(k_\parallel<k_0,v_0=0)$ and $E(|k_x|<k_0,v_0=0)$ are compared. For $E(|k_x|<k_0,v_0=0)$, the DOS is constant and independent on the energy $\omega$. Once the limitation of $|k_x|<k_0$ is changed to be $k_\parallel<k_0$, the DOS $\rho\left(\omega\right)$ is disturbed. Specifically, as $\omega$ moves away from zero energy ($\omega=0$), the DOS $\rho\left(\omega\right)$ decreases rapidly. The largest perturbation occurs at the band edges (i.e., $\omega=\pm\gamma k_0$ or $k_y=\mp k_0$), where vanished DOS is obtained, as highlighted by the magenta circles in Fig. 4(b). The largest perturbation at the band edges indicates that the surface states-mediated RKKY interaction here is not only contributed by the electrons near the Fermi surface (i.e., $\omega=0$ or $k_F=0$) but also by the electrons near the band edges.
\par
Based on the above disturbed DOSs, one can calculate the Green's functions $g(\omega,\pm R_y)$ analytically (see detailed derivation in the Appendix III), which are given by
\begin{eqnarray}
\begin{split}
g\left(\omega,- R_y\right)&=-i\pi^2\left(\frac{-3\gamma v_z}{2vk_0^3}\right)^{1/3}{\rho\left(\omega\right)}^{4/3}e^{-iR_y\omega/\gamma}+O\left(\frac{1}{R^{m>0}}\right), \\
g\left(\omega,R_y\right)&=0-\frac{\omega_-{\rm cos}\left(k_0R_y\right)+\omega_+{\rm sin}\left(k_0R_y\right)}{\left(\omega_+\omega_--2\right)v_z\pi^2\gamma/\left(v\sqrt{\pi k_0}\right)}\frac{1}{R^{5/2}},
\end{split}
\end{eqnarray}
where $\rho(\omega)=2vk_0^3(1-\omega^2/\gamma^2k_0^2)^{3/2}/(-3\pi^2\gamma v_z)$ is the DOS of the surface band and $\omega_{\pm}=\omega/\gamma k_0\pm i$. Compared to the Green's functions of $E(|k_x|<k_0,v_0=0)$ in Eq. (8), one can find that the changed limitation $k_\parallel<k_0$ only modifies the amplitude of $g(\omega,- R_y)$ via the DOS, but it changes the vanished $g(\omega,R_y)$ to be a finite one. Noting that the first terms in the right-hand side of Eq. (11) are obtained by applying the expansion to the integrand of Eq. (6) at the point $k_F=0$ (i.e., Fermi surface), while the second ones come from the expansion at the band edges ($\omega=\pm\gamma k_0$ or $k_y=\mp k_0$). This means that the finite $g(\omega,R_y)$ is completely induced by the electrons near the band edges [magenta circles in Fig. 4(b)], which are allowed to complete a trip from $\mathbf{r}_1$ to $\mathbf{r}_2$, called by us "the band-edge effect".
\par
Plugging the Eq. (11) into the Eq. (7) and integrating out the energy $\omega$, the analytical result of the surface contribution $J$ can be solved as
\begin{eqnarray}
\begin{split}
J\left(R_y\right)=\frac{128v^2k_0^{7/2}C}{3\gamma v_z^2\pi^{1/2}}\frac{\sin\left(k_0R_y\right)-\cos\left(k_0R_y\right)}{R_y^{7/2}}.
\end{split}
\end{eqnarray}
 The above analytical result explains the decay and the oscillation of the RKKY interaction in Fig. 4(a), as denoted by the dashed line.
\par
\begin{figure}[th]
\centering \includegraphics[width=0.48\textwidth]{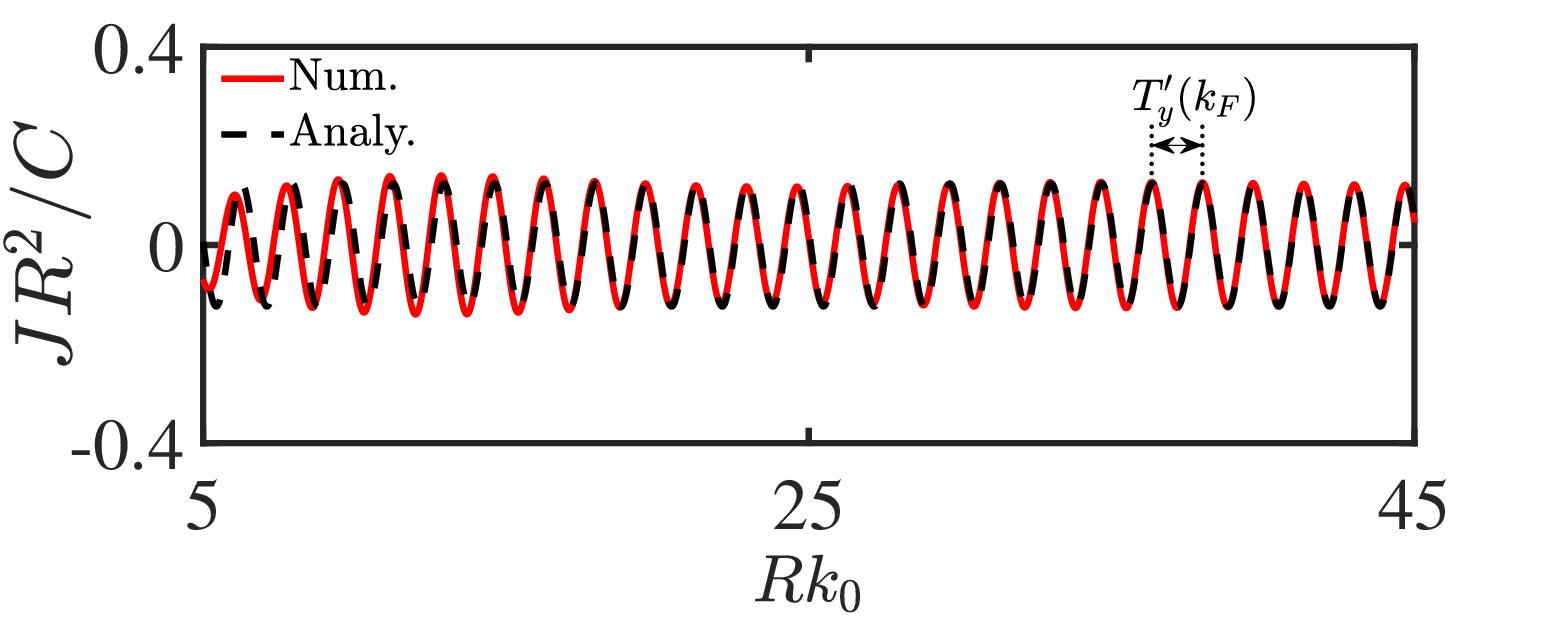}
\caption{(Color online) The $R$-dependent RKKY interaction contributed by $E(|k_x|<k_0,v_0\neq0)$ with impurities placed in $y$ axis. The solid line denotes for the numerical result calculated from Eqs. (6,7) and the dash line refers to the analytical result from Eq. (15).}
\end{figure}

 \begin{figure}[th]
\centering \includegraphics[width=0.45\textwidth]{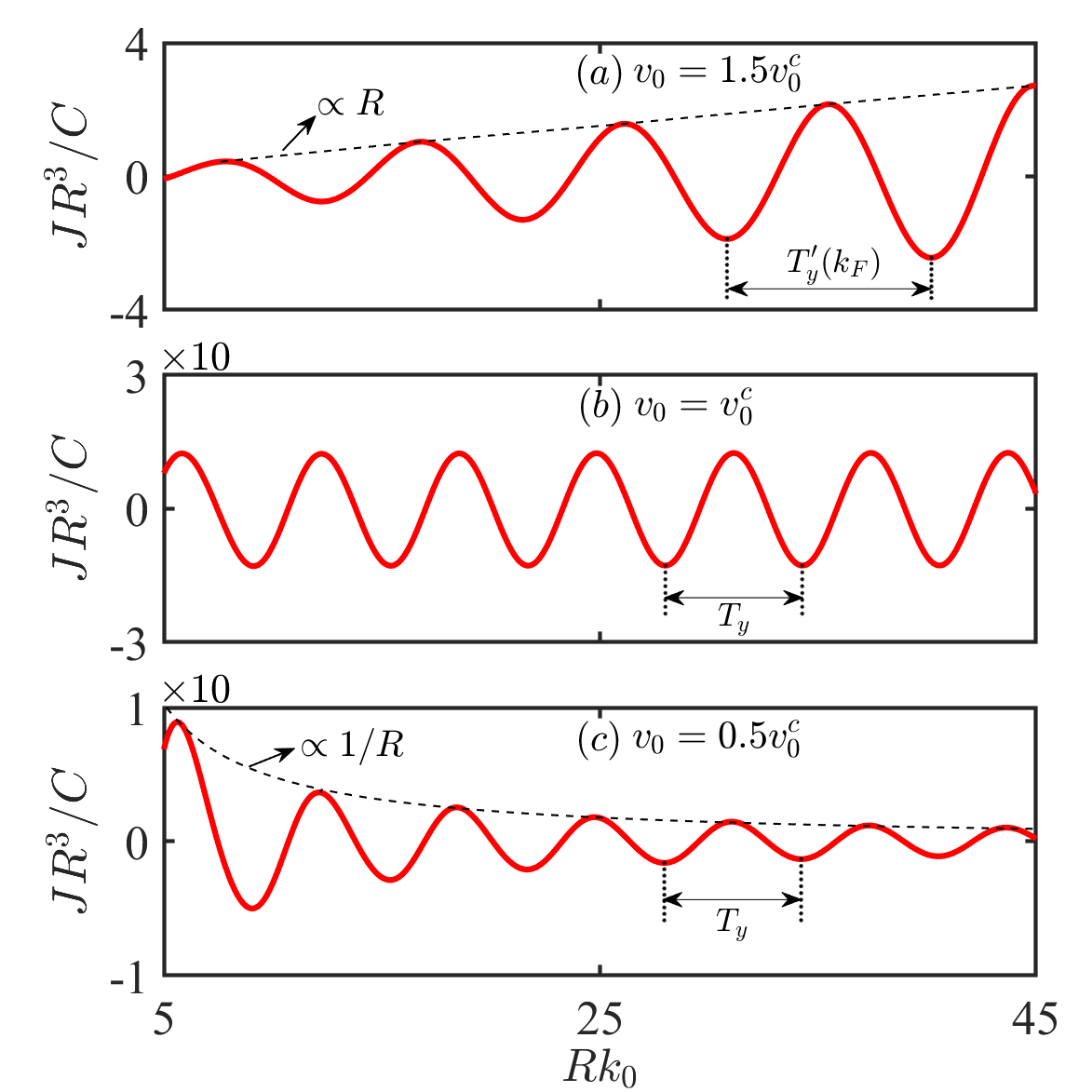}
\caption{(Color online) The $R$-dependent RKKY interaction contributed by $E(k_\parallel<k_0,v_0\neq0)$ with impurities in $y$ axis. Different values of $v_0$ are considered to discuss the competition of the two effects (i.e., the band-edge effect and the bending effect of the surface band).}
\end{figure}

 \begin{figure*}[th]
\centering \includegraphics[width=0.85\textwidth]{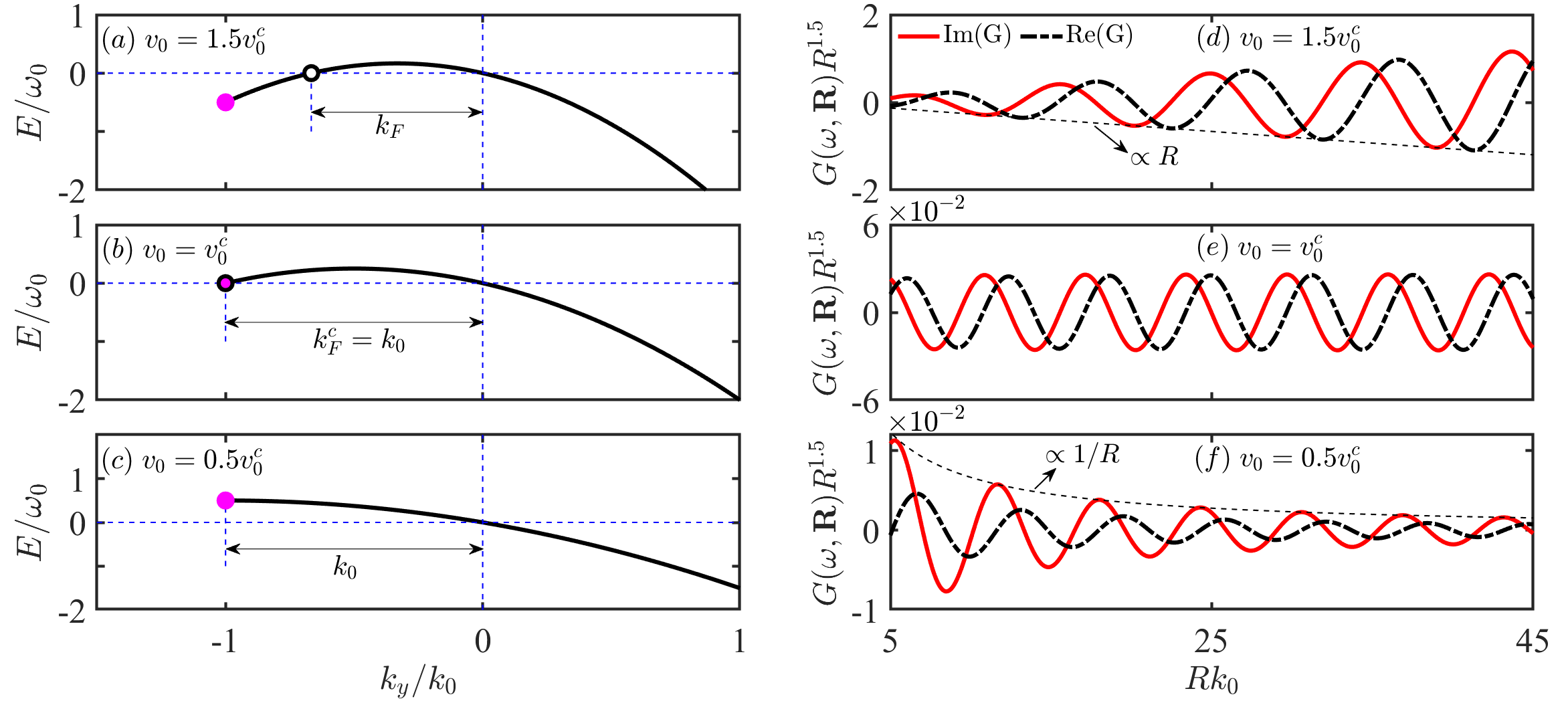}
\caption{(Color online) (a-c) Energy dispersion along $k_y$ axis with different values of $v_0$. The magenta circle denote for the momentum of the band edge, and the hollow circle for the momentum located on the Fermi energy. The Fermi wave numbers are $k_F=\gamma/v_0$, $k^c_F=k_0$ (or $\gamma/v_0$). (d-f) The $R$-dependent Green's function $g\left(\omega,\mathbf{R}\right)$ with different $v_0$. The critical value of $v_0$ is $v_0^c=\gamma/k_0$.}
\end{figure*}
\par
\subsection{RKKY interactions mediated by $E(|k_x|<k_0,v_0\neq0)$}
For the case of $E(|k_x|<k_0,v_0\neq0)$, the numerical results of the surface contribution can also be obtained according to the Eqs. (2) and (6-7), as plotted in Fig. 5. Compared to the case of $E(k_\parallel<k_0,v_0=0)$, the interaction here decays much more slowly with the impurity distance as $1/R_y^2$. In addition, there exists a different oscillation with a period of ${T'_y}=\pi /k_F$, whose Fermi wave number $k_F=\gamma/v_0$ is induced by the bending effect of the surface band. Similar oscillations but induced by the finite Fermi energy ($u_F\neq0$) are also reported in previous literatures\cite{graphene1,graphene2,graphene3,graphene4,graphene5,Weyl1,Weyl2,Weyl3,topo1,surface1,surface2,surface3,surface4,surface5,NLSM,alpha,semiDirac}.
\par
To obtain the analytical result, one have to notice that the dispersion in $k_y$ axis is infinite, i.e., the band-edge effect of $k_y$ axis disappears. Thus, the interaction here is only contributed by the electrons at the Kohn-anomaly point\cite{layer}, which corresponds to
a singular point on the Fermi surface.  According to the properties of singular point, one can use the partial differential equations $\partial k_y/\partial k_x = 0$ [$k_y=\left(-\gamma\pm\sqrt{\gamma^2-4v_0^2k_x^2}\right)/\left(2v_0\right)$] to calculate the positions $(k_{x_i},k_{y_i})$ of the Kohn-anomaly points as
\begin{eqnarray}
\begin{split}
\left(k_{x_1},k_{y_1}\right)&=\left(0,-\frac{\gamma}{v_0}\right), \\
\left(k_{x_2},k_{y_2}\right)&=\left(0,0\right).
\end{split}
\end{eqnarray}
Since the two Kohn-anomaly points are all located on $k_y$ axis (i.e., $k_{x}=0$), one can expand the integrand of Eq. (6) at $k_{x}=0$ to calculate $g\left( \pm R_y,\omega\right)$ (see detailed derivation in the Appendix III) as
\begin{eqnarray}
\begin{split}
g\left(\omega, -R_y\right)&=\frac{\left(1-i\right)k_0^2v\left(\gamma^2-v_0\omega\right)}{2\sqrt{2}v_z\gamma^2\sqrt{\pi v_0\gamma}}\frac{e^{-iR_y\omega/\gamma}}{R_y^{1/2}},  \\
g\left(\omega, R_y\right)&=e^{-i\frac{R_y\gamma}{v_0}}g\left(\omega, -R_y\right).
\end{split}
\end{eqnarray}
Plugging the above Green's functions into Eq. (7) and integrating out the energy $\omega$, one can solve the surface contribution as
\begin{eqnarray}
\begin{split}
J\left(R_y\right)=\frac{8\pi v^2k_0^4C}{v_0v_z^2}\frac{\sin\left(\gamma R_y/v_0\right)}{R_y^{2}}.
\end{split}
\end{eqnarray}
The above analytical result explains the decay and the oscillation of the RKKY interaction in Fig. 5, as denoted by the dashed line.

\subsection{RKKY interaction contributed by the competition
 of the two mechanisms [i.e., $E(k_\parallel<k_0,v_0\neq0)$ ]}
 In this chapter, we would study the effect of the competition
 of the two mechanisms [i.e., $E(k_\parallel<k_0,v_0\neq0)$] on the surface contribution. To facilitate this discussion, the limitation of $k_\parallel<k_0$ is kept unchanged and $v_0$ varies. The numerical results of $J(R_y)$ are shown in Fig. 6. It is found that there exits a critical value $v^c_0=\gamma/k_0$ for $v_0$, which determines the decay and the oscillation of the surface contribution. For $v_0>v_0^c$, the interaction is completely governed by the bending effect while the band-edge effect can be ignored. As a result, the interaction share a same decay law ($1/R_y^2$), as well as a same oscillation [${\rm sin}(\gamma R_y/v_0)$], as that in Eq. (15). Once $v_0$ decreases with $v_0<v_0^c$, the interaction decays fast as $1/R_y^4$, along with an oscillation ${\rm sin}(k_0R_y)$ determined by the band-edge effect. For the critical case $v_0=v^c_0$, an intermediate behavior arises, i.e., $J\propto1/R_y^{3}=1/R_y^{(2+4)/2}$ with the oscillation ${\rm sin}(R_y\gamma/v_0)$ [or  ${\rm sin}(k_0R_y)$].
\par
To understand the above phenomenon, one have to notice that the band-edge effect induced by $k_\parallel<k_0$ only changes the Green's function $g(\omega,R_y)$, while the decay and the oscillation of the other Green's function $g(\omega,-R_y)$ are completely controlled by the bending effect ($v_0\neq0$), which results in $g(\omega,-R_y)\propto e^{-iR_y\omega/\gamma}/R_y^{1/2}$ [Eq. (14)]. Thus, the competition of the two mechanisms is only reflected on $g(\omega,R_y)$. We plot the evolution of $g(\omega,R_y)$ with different $v_0$ in Fig. 7, where the dispersion of the $k_y$ axis is also plotted for a better understanding. As shown in Fig. 7(a), the band edge (the magenta circle) of the surface band is far away from the Fermi energy for $v_0>v_0^c$. Thus, the interaction is mainly contributed by the bending effect, the resulting Fermi wave number $k_F$ (the black circle) naturally leads to $g(\omega,R_y)\propto e^{-ik_FR_y}/R_y^{1/2}$ with $k_F=\gamma/v_0$ [Fig. 7(d) or Eq. (14)]. Differently, for $v_0<v^c_0$, the nonzero Fermi wave number disappears [Fig. 7(c)]. Thus, the bending effect does not work in this case and the electrons near the band edge [the magenta circle in Fig. 7(c)] would play the leading role in contributing $g(\omega,R_y)$, which results in $g(\omega,R_y)\propto {\rm sin}(k_0R_y)/R_y^{5/2}$ [Fig. 7(f) or Eq. (11)]. By integrating out the energy $\omega$ of the product $g(\omega,R_y)g(\omega,-R_y)$, an extra factor $1/R_y$ is generated\cite{pn}. Then, the complex RKKY behaviors arises, i.e., $J(v_0>v_0^c)\propto{\sin}(\gamma R_y/v_0)/R_y^2$ and $J(v_0<v_0^c)\propto{\sin}(k_0 R_y)/R_y^4$ as shown in Fig. 6(a) and 6(c). For the critical case, the two mechanism would operate at the same time to result in an intermediate $g(\omega,R_y)\propto {\rm sin}(k_0R_y)/R_y^{(1/2+5/2)/2}={\rm sin}(k_0R_y)/R_y^{3/2}$, which naturally leads to an intermediate RKKY behavior in Fig. 6(b).
\par
\begin{figure}[th]
\centering \includegraphics[width=0.47\textwidth]{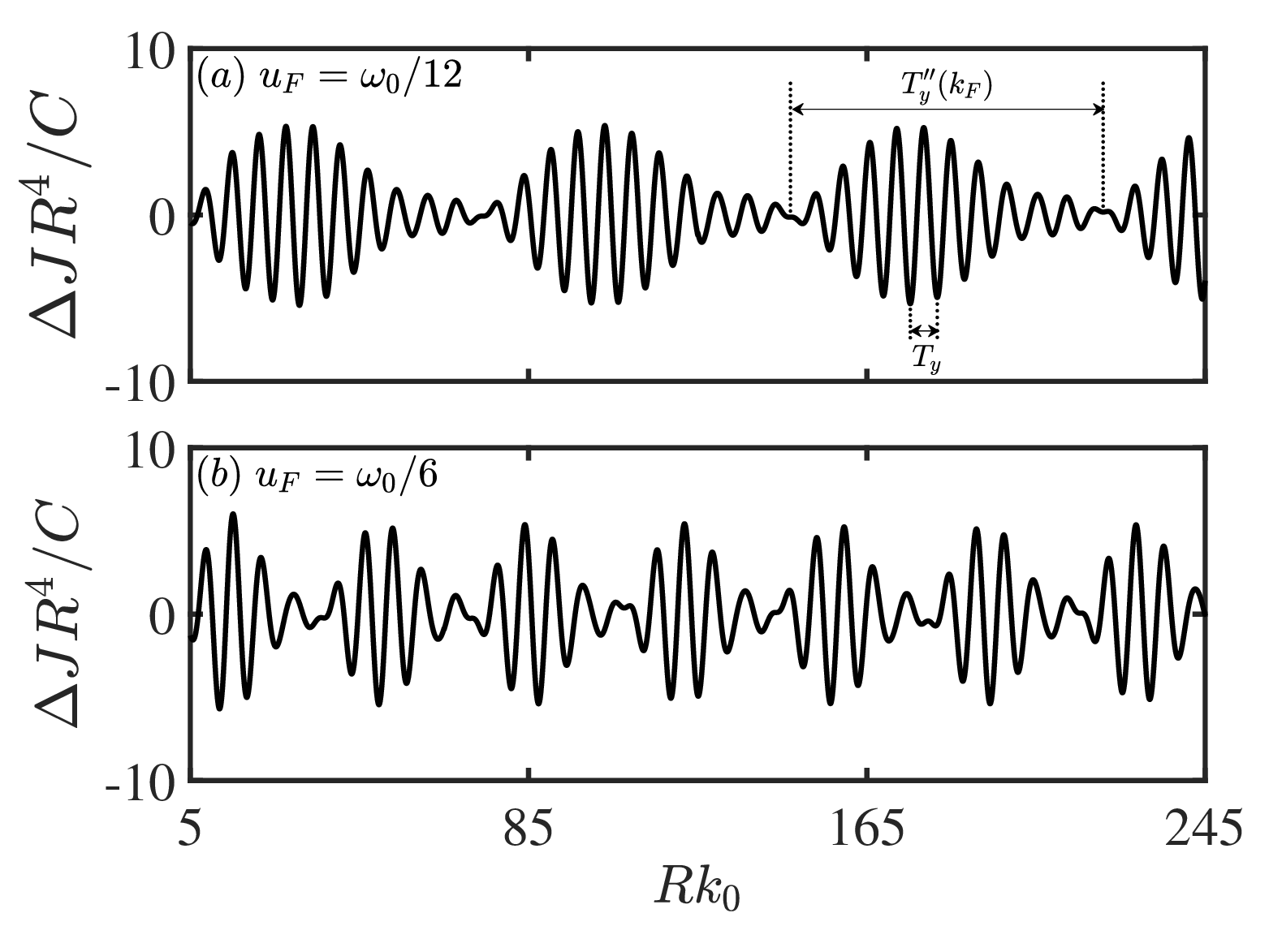}
\caption{(Color online) The $R$-dependent RKKY interaction $\Delta J=J(u_F)-J(0)$ with different finite Fermi energies. The large period of the oscillation is $T^{\prime\prime}_y(k_F)=2\pi/k_F$ with $k_F=(\gamma+\sqrt{4u_Fv_0+\gamma^2})/2 v_0$, and the small one is $T_y=2\pi/k_0$. Here, the surface band is $E(k_\parallel<k_0,v_0\neq0)$ with $v_0=-0.993{\rm eV\AA^2}$\cite{Chan}.}
\end{figure}
Finally, we consider the effect of the finite Fermi energy. To highlight the oscillating characteristic of the magnetic interaction, the interaction by subtracting the case of zero Fermi energy, i.e, $J(u_F)-J(0)$, is plotted in Fig. 8. Compared to the case of $u_F=0$, the main difference here is that the original one-period oscillation is changed to be an oscillation with two periods, i.e., exhibiting an interesting battering pattern. The small period $T_y=2\pi/k_0$ is originated from the band-edge effect, while the large one is induced by the finite Fermi energy $u_F$, which results in $T_y^{\prime\prime}(k_F)=2\pi/k_F$ with the Fermi wave number $k_F=(\gamma+\sqrt{4u_Fv_0+\gamma^2})/2 v_0$. Similar battering pattern is also obtained in bulk contributions of WSMs/DSMs\cite{Weyl1,Weyl2,surface4,surface5} but with different physics, which is attributed to the combined effect of finite $u_F$ and the splitting of the Weyl/Dirac points.
\par
Overall, our results exhibited in Sec. III (${\rm B}$-${\rm D}$) suggest that the oscillations and the decays of the RKKY interaction are highly sensitive to the shape of the surface band of WSMs. Under the peculiar mechanisms, these RKKY behaviors are unique and significantly different from that of other 2D topological surface bands (e.g., the helical surface states of topological insulators\cite{topo1}). Thus, the interaction here can be used to characterize the dispersive nature of the surface states of WSMs.

\section{Summary}
We have explored the RKKY interaction mediated by a single surface band in WSMs. The nonzero surface contribution can be induced by two mechanisms, i.e., either by the band-edge effect or by the bending effect of the surface band. The cases here are significantly different from that of previous literatures, where surface states should couple with bulk states\cite{surface4} or other surface states of different spin\cite{surface5} to result in nonzero interaction. For impurities deposited in the direction perpendicular to the Weyl points splitting, the surface contributions here always decay much more slowly with impurity distance than that of bulk contribution. Under different mechanisms, the surface contribution exhibits different decay laws and oscillations. Moreover, these two mechanisms would compete with each other to result in more complex RKKY behaviors. In addition, an interesting oscillation with peculiar battering pattern is obtained due to the combined effect of the band-edge effect and the finite Fermi energy. Due to the sensitivity of the RKKY interaction to the shape of the surface band of WSMs, magnetic signals (peculiar oscillations and slowly decaying laws) can be extracted to characterize the dispersive nature of surface states of WSMs. Compared to the surface bands in previous literatures, our model used here is more practical, and is conducive to the understanding of the surface magnetic properties in WSMs.

\acknowledgements This work was supported by the National Natural Science Foundation of China (Grant Nos. 12104167, 12174121, 11904107, 11774100), by Guangdong Basic and Applied Basic Research Foundation (Grant No. 2020A1515111035), by GDUPS (2017) and by Key Program for Guangdong NSF of China (Grant No. 2017B030311003).
\newpage
\begin{widetext}
\appendix
\renewcommand\thesection{\Roman{section}}
\renewcommand\thesubsection{\Alph{subsection}}
\def\CTeXPreproc{Created by ctex v0.2.12, don't edit!}
\numberwithin{equation}{section}
\section{Phase transition from NLSMs to WSMs by the circularly polarized light}
The model employed in Eq. (1) can be realized by considering the effect of a periodic driving to the following model of NLSMs,
\begin{equation}
H_{{\rm NLSM}}=v\left(k_\parallel^2-\kappa_0^2\right)\tau_z+v_zk_z\tau_y+v_0\left(k_\parallel^2-\kappa_0^2\right)\tau_0,
\end{equation}
which is extracted from the ${\rm Ca_3P_2}$-like materials\cite{Chan}. For the sake of concreteness, a beam of circularly polarized light is assumed to be injected in the $x$ axis. The corresponding vector potential is described as $\mathbf{A}(t)=A_0[0,\cos(\omega t),\sin(\omega t)]$ with period $T=2\pi/\omega$.  By applying the Peierls substitution $\mathbf{k}\rightarrow \mathbf{k}+e\mathbf{A}/\hbar$, the system Hamiltonian becomes time-dependent. Using the Floquet theory\cite{floquet3} with the off-resonant condition of $A^{2}/\omega \gg 1$, the modified part of the Hamiltonian induced by
light reads as
\begin{equation}
H'_{{\rm NLSM}}=V_{0}+\sum_{n\geq 1}\frac{\left[ V_{+n},V_{-n}\right] }{\hbar
\omega}+O\left(\frac{1}{\omega ^{2}}\right),
\end{equation}
where $V_{n}=\frac{1}{T}%
\int_{0}^{T}H(t)e^{-in\hbar \omega t}$ is solved as
\begin{eqnarray}
\begin{split}
V_0&=H_{{\rm NLSM}}+\frac{vk_A^2}{2}\tau_z+\frac{v_0k_A^2}{2}\tau_0,\\
V_{\pm1}&=k_A\left(v_0k_y\tau_0+v k_y\tau_z\pm i\frac{v_z}{2}\tau_y\right),\\
V_{\pm2}&=\frac{k_A^2}{4}\left(v_0\tau_0+v\tau_z\right),
\end{split}
\end{eqnarray}
and $V_n=0$ for $|n|>2$. Here, we set $k_A=eA_0/\hbar$ for simplicity. According to the above equations, one can obtain the following effective
Hamiltonian as
\begin{equation}
H'_{{\rm NLSM}}=H_{{\rm NLSM}}+\frac{vk_A^2}{2}\tau_z+\frac{v_0k_A^2}{2}\tau_0-\frac{2vv_zk_A^2}{\omega}k_y\tau_x.
\end{equation}
By applying the parameter transformation $(\kappa^2_0-k_A^2/2,-2vv_zk_A^2/\omega)=(k^2_0,\gamma)$ and dropping the constant term $-v_0k_0^2\tau_0$, one can obtain the same Hamiltonian as that in Eq. (1).

\section{The numerical results of the RKKY interaction with impurities in $x$ axis}
\begin{figure}[th]
\centering \includegraphics[width=0.45\textwidth]{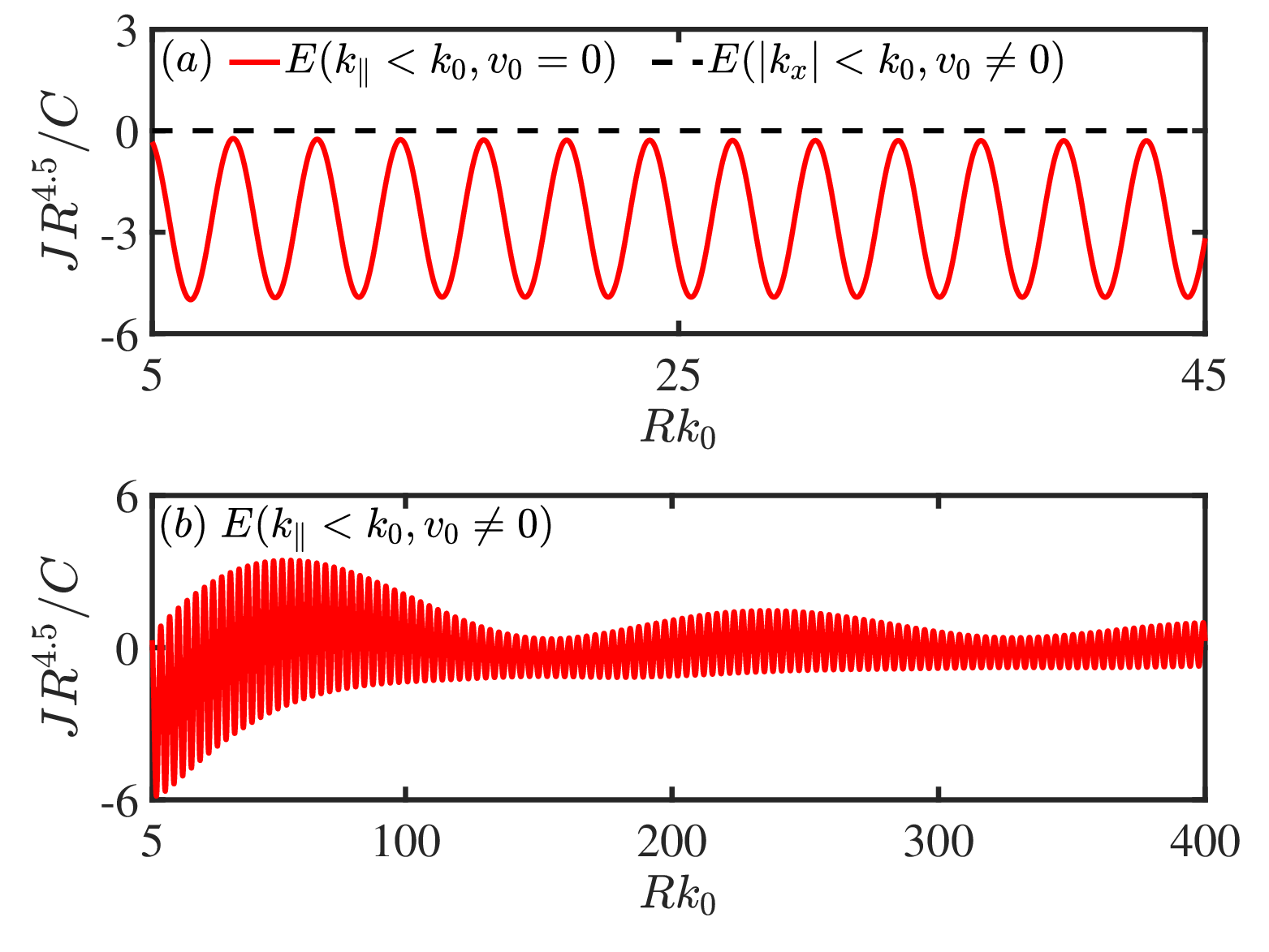}
\caption{(Color online) The $R$-dependent RKKY interaction contributed by (a) $E(k_\parallel<k_0,v_0=0)$, $E(|k_x|<k_0,v_0\neq0)$ and (b) $E(k_\parallel<k_0,v_0\neq0)$, impurities are placed in the $x$ axis. $C=\lambda^2/(2\pi)^3$ and other parameters are the same as that in Fig. 3(c) of the main text.}
\end{figure}
For the case of $E(k_\parallel<k_0,v_0=0)$, according to the Eqs. (6-7) of the main text, the numerical results of the interaction $J$ with impurities in the $x$ axis can be calculated, as shown by the solid line in Fig. 9(a). Here, the interaction $J$ decays with $R$ as $R^{9/2}$, along with an oscillation whose period is $T_x=\pi/k_F$ ($k_F=k_0$). This oscillation is different from the case of $y$ axis since it is induced by the Fermi wave number $k_F$, which characterizes the projection of the edge of the Fermi surface in $k_x$ axis. The oscillation of the same period can also be obtained by the bulk contribution but with its mechanism attributed to the splitting of the Weyl points\cite{Weyl1,Weyl2}. Similarly, the interaction $J(R_x)$ of $E(|k_x|<k_0,v_0\neq0)$ can also be calculated numerically, as shown by the dashed line in Fig. 9(a). It is found that $J$ always vanishes, which echoes with the case of the trivial phase factor $\vartheta$ ($\vartheta=\pi$) in Fig. 3(d) of the main text. For the case of $E(k_\parallel<k_0,v_0\neq0)$, the numerical result of $J(R_x)$ is shown in Fig. 10. Compared to the case of $E(k_\parallel<k_0,v_0=0)$, it decays much more fast with $R$.
\par
The interaction here is insignificant since it exhibits a faster decaying law as compared to the case with impurities in $y$ axis. Thus, we only focus on $J(R_y)$ in the main text.

\section{Derivation of the analytical RKKY interactions}
In this section, we would show the detailed derivation of the analytical RKKY interactions. Two cases with different dispersions $E(k_\parallel<k_0,v_0=0)$ and $E(|k_x|<k_0,v_0\neq0)$ would be studied respectively. For zero Fermi energy $u_F=0$, the amplitude of the RKKY interaction $J$ of Eq. (7) in the main text can be written as
\begin{equation}
J=-\frac{8\lambda^2 }{\pi }\mathrm{Im}\int_{-\infty }^{0} g\left(
\mathbf{R},\omega \right)  g\left( -\mathbf{R},\omega \right) d\omega ,
\end{equation}
with
\begin{eqnarray}
\begin{split}
g\left( \pm \mathbf{R},\omega\right)&=\frac{1}{\left(2\pi\right)^2}\int_{-k_0}^{k_0} dk_y  \int_{-\sqrt{k_0^2-k_y^2}}^{\sqrt{k_0^2-k_y^2}} dk_x\frac{v}{v_z} \frac{k_0^2-k_x^2-k_y^2}{\omega -\gamma k_y+i\eta}e^{\pm i\mathbf{kR}} \;\text{for}\;E(k_\parallel<k_0,v_0=0) ,\\
g\left( \pm \mathbf{R},\omega\right)&=\frac{1}{\left(2\pi\right)^2}\int_{-k_0}^{k_0} dk_x\int_{-\infty}^{\infty}dk_y\frac{v}{v_z} \frac{k_0^2-k_x^2}{\omega -v_0k_\parallel^2-\gamma k_y+i\eta}e^{\pm i\mathbf{kR}}\;\text{for}\;E(|k_x|<k_0,v_0\neq0),\\
\end{split}
\end{eqnarray}
where $\eta$ is a positive infinitesimal.

\numberwithin{equation}{subsection}
\subsection{The case of $E(k_\parallel<k_0,v_0=0)$}
According to Eqs. (III.1) and (III.2), for impurities placed along $y$ axis (i.e., $R_x=0$ and $R_y\neq0$), $g\left( \omega ,\pm R_y\right)$ can be solved as
\begin{eqnarray}
\begin{split}
g\left( \pm R_y,\omega\right)&=\frac{1}{\left(2\pi\right)^2} \int_{-k_0}^{k_0} dk_y  \int_{-\sqrt{k_0^2-k_y^2}}^{\sqrt{k_0^2-k_y^2}} dk_x\frac{v}{v_z} \frac{k_0^2-k_x^2-k_y^2}{\omega-\gamma k_y+i\eta}e^{\pm i k_yR_y}, \\
&=\frac{4v}{3v_z\left(2\pi\right)^2} \int_{-k_0}^{k_0} dk_y   \frac{\left(k_0^2-k_y^2\right)^{3/2}}{\omega-\gamma k_y+i\eta}e^{\pm i k_yR_y},\\
&=\frac{vk_0^3}{-3\gamma v_z\pi^2} \int_{-\pi/2}^{\pi/2} dx   \frac{{\rm cos}^4\left(x\right)}{ {\rm sin}\left(x\right)-\left(\omega+i\eta\right)/\gamma k_0 }e^{\pm i k_0R_y{\rm sin}\left(x\right)}, \\
&=f\left( \omega ,\pm R_y\right)+\frac{2}{k_0^2}\frac{d^2f\left( \omega ,\pm R_y\right)}{R_y^2}+\frac{1}{k_0^4}\frac{d^4f\left( \omega ,\pm R_y\right)}{R_y^4},
\end{split}
\end{eqnarray}
with
\begin{eqnarray}
\begin{split}
f\left( \pm R_y,\omega\right)=\frac{vk_0^3}{-3\gamma v_z\pi^2} \int_{-\pi/2}^{\pi/2} dx   \frac{1}{ {\rm sin}\left(x\right)-\left(\omega+i\eta\right)/\gamma k_0 }e^{\pm i k_0R_y{\rm sin}\left(x\right)},
\end{split}
\end{eqnarray}
where a parameter transformation $k_y=k_0{\rm sin}\left(x\right)$ is used. As addressed in Sec. III-B of the main text, the surface contribution here is not only contributed by the electrons near the Fermi energy (i.e., $k_y=0$ or $x=0$) but also contributed by the electrons near the band edge (i.e., $k_y=\pm k_0$ or  $x=\pm \pi/2$). Thus, $f\left( \omega ,\pm R_y\right)$ can be expressed as
\begin{eqnarray}
\begin{split}
f\left( \pm R_y,\omega\right)=f_0\left( \omega ,\pm R_y\right)+f_+\left( \omega ,\pm R_y\right)+f_-\left( \omega ,\pm R_y\right),
\end{split}
\end{eqnarray}
 where $f_n\left( \omega ,\pm R_y\right)$ refers to the approximate result of $f\left( \omega ,\pm R_y\right)$ by considering the integrand of Eq. (A.2) around the points $n\pi/2$ with ($n=0,+,-$).

 \par
 Specifically, around the point $x=0$, $f_0\left( \omega ,\pm R_y\right)$ can be approximated as
 \begin{eqnarray}
\begin{split}
f_0\left( \pm R_y,\omega\right)=\frac{vk_0^3}{-3v_z\gamma\pi^2}\int_{-\infty}^{\infty}dx\frac{1}{x-\omega/\gamma k_0}e^{\pm ik_0R_yx}.
\end{split}
\end{eqnarray}
After some algebraic calculations, $f_0\left( \omega ,\pm R_y\right)$ can be solved as
 \begin{eqnarray}
\begin{split}
f_0\left(-R_y,\omega\right)&=\frac{i2vk_0^3}{3\pi v_z\gamma}e^{-iR_y\omega/\gamma}, \\
f_0\left( R_y,\omega\right)&=0.
\end{split}
\end{eqnarray}
Around the point $x=\pi/2$, $f_+\left( \omega ,\pm R_y\right)$ can be approximated as
 \begin{eqnarray}
\begin{split}
f_+\left( \pm R_y,\omega\right)&=\frac{vk_0^3}{-3v_z\gamma\pi^2}\int _{-\infty}^{0}dx' \frac{1}{1-\omega/\gamma k_0}e^{\pm i\left(1-{x'}^2/2\right)k_0R_y},
\end{split}
\end{eqnarray}
where a parameter transformation $x=x^\prime+\pi/2$ is used. Then, $f_+\left( \pm R_y,\omega\right)$ can be easily solved as
 \begin{eqnarray}
\begin{split}
f_+\left(\pm R_y,\omega\right)=\frac{vk_0^{5/2}\left(1\mp i\right)}{6\pi^{3/2}\gamma v_z\left(\omega/\gamma k_0-1\right)}\frac{1}{R_y^{1/2}}e^{\pm ik_0R_y}.
\end{split}
\end{eqnarray}
Similarly, by applying a parameter transformation $x=x^{\prime\prime}-\pi/2$ to $f_-\left( \pm R_y,\omega\right)$, the approximate result of $f_-\left( \pm R_y,\omega\right)$ at the point $x=-\pi/2$ can be solved as
 \begin{eqnarray}
\begin{split}
f_-\left(\pm R_y,\omega\right)&=\frac{vk_0^3}{3v_z\gamma\pi^2}\int ^{\infty}_{0}dx'' \frac{1}{1+\omega/\gamma k_0}e^{\mp i\left(1-{x''}^2/2\right)k_0R_y}, \\
&=\frac{vk_0^{5/2}\left(1\pm i\right)}{6\pi^{3/2}\gamma v_z\left(\omega/\gamma k_0+1\right)}\frac{1}{R_y^{1/2}}e^{\mp ik_0R_y}.
\end{split}
\end{eqnarray}
Plugging Eqs. (A.5), (A.7) and (A.8) into Eq. (A.3) and Eq. (A.1), the Green's function $g\left( \pm R_y,\omega\right)$ can be solved as
\begin{eqnarray}
\begin{split}
g\left(- R_y,\omega\right)&=-i\pi^2\left(\frac{-3\gamma v_z}{2vk_0^3}\right)^{1/3}{\rho\left(\omega\right)}^{4/3}e^{-iR_y\omega/\gamma}+O\left(\frac{1}{R^{m>0}}\right), \\
g\left(R_y,\omega\right)&=0-\frac{\omega_-{\rm cos}\left(k_0R_y\right)+\omega_+{\rm sin}\left(k_0R_y\right)}{\left(\omega_+\omega_--2\right)v_z\pi^2\gamma/\left(v\sqrt{\pi k_0}\right)}\frac{1}{R^{5/2}},
\end{split}
\end{eqnarray}
where $\rho(\omega)=2vk_0^3(1-\omega^2/\gamma^2k_0^2)^{3/2}/(-3\pi^2\gamma v_z)$ is the DOS of surface band and $\omega_{+}=\omega/\gamma k_0\pm i$. The first terms in the right-hand side of the above equations are contributed by the electrons near the Fermi energy (i.e., $k_y=0$). The second terms are induced by the electrons near the band edge ($k_y=\pm k_0$). Plugging the above equations into Eq. (III.1) and integrating out the energy $\omega$, one can obtain the RKKY interaction
\begin{eqnarray}
\begin{split}
J\left(R_y\right)=\frac{16v^2k_0^{7/2}\lambda^2}{3\gamma v_z^2\pi^{7/2}}\frac{\sin\left(k_0R_y\right)-\cos\left(k_0R_y\right)}{R_y^{7/2}}.
\end{split}
\end{eqnarray}

\subsection{The case of $E(|k_x|<k_0,v_0\neq0)$}
Here, we focus on the case of $E(|k_x|<k_0,v_0\neq0)$ with impurities placed along $y$ axis. According to Eq. (III.2), the Green's function $g\left( \pm R_y,\omega\right)$ can be rewritten and solved as
\begin{eqnarray}
\begin{split}
g\left( \pm R_y,\omega\right)&=\frac{1}{\left(2\pi\right)^2}\int_{-k_0}^{k_0} dk_x\int_{-\infty}^{\infty}dk_y\frac{v}{v_z} \frac{k_0^2-k_x^2}{\omega -v_0k_\parallel^2-\gamma k_y+i\eta}e^{\pm ik_yR_y},\\
&=\frac{-iv}{\pi v_z}e^{\mp\frac{i\gamma R_y}{2v_0}}\int_0^{k_0}dk_x\frac{k_0^2-k_x^2}{\sqrt{\gamma^2+4v_0\left(\omega+i\eta-v_0k_x^2\right)}}e^{i\frac{R_y\sqrt{\gamma^2+4v_0\left(\omega+i\eta-v_0k_x^2\right)}}{2v_0}}.\\
\end{split}
\end{eqnarray}
Noting that the RKKY interaction here is mainly determined by the electrons at the Kohn-anomaly point, which corresponds to a singular point on the Fermi surface. According to the properties of singular point, one can use the partial differential equations $\partial k_y/\partial k_x = 0$ [$k_y=\left(-\gamma\pm\sqrt{\gamma^2-4v_0^2k_x^2}\right)/\left(2v_0\right)$] to calculate the positions $(k_{x_i},k_{y_i})$ of the Kohn-anomaly points as
\begin{eqnarray}
\begin{split}
\end{split}
\left(k_{x_1},k_{y_1}\right)&=\left(0,-\frac{\gamma}{v_0}\right), \\
\left(k_{x_2},k_{y_2}\right)&=\left(0,0\right).
\end{eqnarray}
Noting that the two Kohn-anomaly points are all located on $k_y$ axis (i.e., $k_x=0$), thus one can expanding the integrand of Eq. (B.1) at $k_x=0$ to calculate $g\left( \pm R_y,\omega\right)$ as
\begin{eqnarray}
\begin{split}
g\left(  \pm R_y,\omega\right)&=\frac{-iv}{\pi v_z}e^{\mp\frac{i\gamma R_y}{2v_0}}\int_0^{\infty}dk_x\frac{k_0^2-k_x^2}{\sqrt{\gamma^2+4v_0\omega}}e^{iR_y\left(\frac{\sqrt{\gamma^2+4v_0\omega}}{2v_0}-\frac{v_0k_x^2}{\sqrt{\gamma^2+4v_0\omega}}\right)},\\
&=\frac{1}{\gamma^2+4v_0\omega}\frac{\left(\sqrt{\gamma^2+4v_0\omega}-i2v_0k_0^2R_y\right)\left(i+\sqrt{\gamma^2+4v_0\omega}\sqrt{\frac{1}{\gamma^2+4v_0\omega}}\right)}{4\sqrt{2\pi}v_0v_z\sqrt{-v_0R_y^3}/v}e^{iR_y\left(\frac{\sqrt{\gamma^2+4v_0\omega}}{2v_0}\mp\gamma\right)}.
\end{split}
\end{eqnarray}
Expanding the result of $g\left(  \pm R_y,\omega\right)$ at the energy of $\omega=0$ and keeping it to the lower order term of $\omega$,  $g\left(  \pm R_y,\omega\right)$ can be further simplified as
\begin{eqnarray}
\begin{split}
g\left( \pm R_y,\omega\right)&=-\left(\frac{1}{\gamma^2}\right)^{1/4}\frac{v\left(1+i\right)\left(v_0\omega-\gamma^2\right)\left(2v_0\omega+\gamma^2+i2v_0k_0^2\gamma R_y\right)}{4\sqrt{2\pi}v_z\gamma^3\left(-v_0R_y\right)^{3/2}}e^{-\frac{iR_y\left(\gamma^2\pm\gamma^2+2v_0\omega\right)}{2v_0\gamma}}.
\end{split}
\end{eqnarray}
Plugging the above equation into Eq. (III.1) and integrating out the energy $\omega$, one can obtain the RKKY interaction
\begin{eqnarray}
\begin{split}
J\left(R_y\right)=\frac{v^2k_0^4\lambda^2}{\pi^2v_0v_z^2}\frac{\sin\left(\gamma R_y/v_0\right)}{R_y^{2}}.
\end{split}
\end{eqnarray}
\end{widetext}

\end{document}